\documentclass[twocolumn,prb,aps,superscriptaddress,showpacs]{revtex4}
\usepackage{graphicx}
\begin{document}
\title {Effect of a tilted magnetic field on the orientation of Wigner crystals}
\author{Shi-Jie Yang}
\affiliation{Department of Physics, Beijing Normal University,
Beijing 100875, China}

\begin{abstract}
We study the effect of a tilted magnetic field on the orientation
of Wigner crystals by taking account of the width of a quantum
well in the $z$-direction. It is found that the cohesive energy of
the electronic crystal is always lower for the $[110]$ direction
parallel to the in-plane field. In a realistic sample, a domain
structure forms in the electronic solid and each domain orients
randomly when the magnetic field is normal to the quantum well. As
the field is tilted an angle, the electronic crystal favors to
align along a preferred direction which is determined by the
in-plane magnetic field. The orientation stabilization is
strengthened for wider quantum wells as well as for larger tilted
angles. Possible consequence of the tilted field on the transport
property in the electronic solid is discussed.

\end{abstract}
 \maketitle
\section{introduction}
It was initially predicted by Wigner that two-dimensional (2D)
electrons crystallize into a triangular lattice in the low density
limit where the electron-electron interactions dominate over the
kinetic energy. In an ideally clean 2D system, the critical $r_s$
($r_s=U/\epsilon_F$, corresponding to the ratio of the Coulomb
energy scale U to the kinetic energy scale of the Fermi energy
$\epsilon_F$) was presented to be $37\pm 5$ from quantum Monte
Carlo simulations \cite{Tanatar}. A strong magnetic field
perpendicular to the 2D plane can effectively localize electron
wave functions while keeping the kinetic energy
controlled\cite{Yosh}. Since this lessens the otherwise severe
low-density condition, it is believed that the Wigner crystal (WC)
can be stabilized in a sufficiently strong magnetic
field\cite{Lam,Kuk,Santos}. Approximate calculations\cite{CM1}
have shown that the WC becomes the lowest energy state when the
filling factor $\nu < 1/6$ for the $GaAs/AlGaAs$ electron system
and around $\nu=1/3$ for the hole system. Since the impurities pin
the electronic crystal, a domain structure forms in a realistic
sample\cite{Sher}. While the electrons in a domain have an order
as they are in the ideal crystal, the orientations of the domains
are random.

Presently, the measurement in a tilted field has become an established technique to explore the various correlated
properties in single layer as well as in double layers 2D electron systems \cite{Cha1}. In a previous
work\cite{Yu}, we have compared the ground state energies of the generalized Laughlin liquid \cite{yang} to the
electronic solid state at a given tilted angle. It was found that the critical filling factor $\nu_c$ at the
solid-liquid transition increases with increasing tilted angle.

In this work, we will examine the relation of the orientation of
the hexagonal WC with the in-plane magnetic field as well as the
width of the 2D quantum well. In a wide quantum well, the electron
wave function extends in the $z$-direction, hence may reduce the
coulomb interactions. The in-plane field deforms the electron wave
function, causing the interaction energy to vary according to the
different patterns or orientations of the electronic crystals. We
calculate the cohesive energy of the electronic crystal in a
Hartree-Fock (HF) approximation. We find that it becomes
anisotropic in the tilted magnetic field. The $[110]$ axis of the
hexagonal electronic crystal favors to align along the direction
of the in-plane magnetic field. This trend of orientation
stabilization is strengthened for larger tilted angles. It also
shows that the energy difference between two orthogonal
orientations of the electronic crystal increases with the width of
the quantum well. The in-plane field favors the domains to orient
to the same direction. Thus, the effective impurity density is
reduced as the field is tilted. We will discuss the possible
consequence of such effect on the transport properties in the
electronic solids.

\section{anisotropic cohesive energy of the electronic crystal}
Consider an electron moving on a $x$-$y$ plane under the influence
of a strong magnetic field which is tilted an angle $\theta$ to
the normal, with $\vec{B}=(B\tan\theta,0,B)$. The electron is
confined in a harmonic potential $V(z)={1\over 2}m_b \Omega^2 z^2$
in the $z$-direction, where $m_b$ is the band mass of the electron
and $\Omega$ the characteristic frequency. Such a quantum well has
been chosen to deal with many quantum Hall
systems\cite{Phillips,Jungwirth} to substitute the realistic
potential which is either triangular or square. It was also used
to discuss the giant magneto-resistance induced by a parallel
magnetic field \cite{das}. We work in the "Landau gauge" by
choosing the vector potential $\vec{A}=\{0, x B_z-z B_x, 0\}$. The
single particle wave function for the lowest LL are:
\begin{eqnarray}
\phi_{X}(\vec{r})&=&{1\over \sqrt{L_y}}e^{-iXy/l_B^2}
    \Phi_0^{\omega_+}\left({-(x-X)\sin\tilde{\theta}+z\cos\tilde{\theta}\over l_{+}}\right) \nonumber\\
   & & \times \Phi_0^{\omega_-}\left({(x-X)\cos\tilde{\theta}+z\sin\tilde{\theta}\over l_{-}}\right),
\label{one}
\end{eqnarray}
where $l_B$ is the magnetic length and $l_\pm^2=\hbar/m\omega_\pm$. $X$ is an integer multiple of $2\pi
l_B^2/L_y$. $\Phi_0^{\omega_\pm}$ is the harmonic oscillator wave function in the lowest energy level
corresponding to the frequencies $\omega_\pm$ and
$\tan\tilde{\theta}=\frac{\omega_c^2}{\omega_+^2-\omega_c^2}\tan\theta$, with the cyclotron frequency
$\omega_c=eB/m_bc$. The frequencies $\omega_\pm$ are given by\cite{yang}
\begin{equation}
\omega_\pm^2={1\over 2}(\Omega^2+\frac{\omega_c^2}{\cos^2 \theta})
\pm\sqrt{{1\over 4}(\Omega^2-\frac{\omega_c^2}{\cos^2 \theta})^2
+\Omega^2 \omega_c^2 \tan^2\theta}.
\end{equation}

The Hamiltonian is given by
\begin{equation}
\hat{H}=\frac{1}{2L_{x}L_{y}} \sum\limits_{\vec{q}}
      \hat{\rho}(\vec{q})v(\vec{q})\hat{\rho}(-\vec{q}),
\label{ham}
\end{equation}
where $v(\vec{q})=\frac{4\pi e^2}{\kappa_0 (\vec{q}_{\Vert}^2+q_z^2)}$ is the Fourier transformation of the
Coulomb interaction. Here $\vec{q}_{\Vert}$ is the in-plane momentum and $q_z$ is the momentum perpendicular to
the quantum well.

From Eq.(\ref{one}), the electron density operator is expressed in the momentum space as
\begin{equation}
\hat{\rho}(\vec{q})=\sum_{X}e^{iq_xX} a_{X_-}^\dagger a_{X_+}F^{\theta}(\vec{q}), \label{density}
\end{equation}
where $X_{\pm}=X\pm q_yl_B^2/2 $. $a_{X}^\dagger (a_{X})$ creates (destroys) an electron in the state $\phi_{X}$.
Here $F^\theta(\vec{q})=e^{-\gamma^2/4-\alpha^2/4}$, with
\begin{eqnarray}
\alpha^2&=&(q_x\cos\tilde{\theta}-q_z\sin\tilde{\theta})^2 l_-^2
         +q_y^2l_B^4/l_-^2\cos^2\tilde{\theta}  \nonumber \\
\gamma^2&=&(q_z\cos\tilde{\theta}+q_x\sin\tilde{\theta})^2 l_+^2
         +q_y^2l_B^4/l_+^2\sin^2\tilde{\theta}.
\end{eqnarray}

Substitute Eq.(\ref{density}) into Eq.(\ref{ham}) and carry out the usual procedure of the HF decoupling of the
Hamiltonian, we get
\begin{equation}
H_{HF}=\frac{n_{L}}{2} \sum\limits_{\vec{q}_{\Vert}}u_{HF}(\vec{q}_{\Vert})
      \Delta(-{\vec{q}_{\Vert}})\sum\limits_{X}e^{-iq_{x}X}
      a^{\dag}_{X_{+}}a_{X_{-}},
\end{equation}
where $n_{L}=1/2\pi l_B^{2}$ is the density of one completely filled LL and
\begin{equation}
\Delta(\vec{q}_{\Vert})=\frac{2\pi l_B^{2}}{L_{x}L_{y}}\sum
      \limits_{X} e^{-iq_{x}X}\langle a^{\dag}_{X_{+}}
      a_{X_{-}} \rangle
\end{equation}
is the order parameter of the charge density wave (CDW). The HF potential is denoted with
$u_{HF}(\vec{q}_{\Vert})=u_{H}(\vec{q}_{\Vert})-u_{ex}(\vec{q}_{\Vert})$. The Hartree term
$u_{H}(\vec{q}_{\Vert})$ is given by (in units of $e^2/\kappa_0 l_B$)
\begin{equation}
u_H(\vec{q}_{\Vert})=\int\frac{dq_z}{\pi l_B}\frac{1}{\vec{q}^2_\Vert+q^2_z} [F^\theta(\vec{q})]^2,
\end{equation}
and the exchange term $u_{ex}(\vec{q}_{\Vert})$ in the reciprocal
space turns out to be proportional to the real-space Hartree
potential as,\cite{Phillips,Koulakov}
\begin{equation}
u_{ex}(\vec{q}_{\Vert})=-2\pi l_B^2\int \frac{d\vec{p}_{\Vert}}{(2\pi)^2}
u_H(\vec{p}_{\Vert})e^{i\vec{p}_{\Vert}\times \vec{q}_{\Vert}l_B^2}.
\end{equation}

Allowing the charge density wave by making ansatz in the plane
\begin{equation}
<a^\dagger_{X-Q_y l_B^2/2}a_{X+Q_y l_B^2/2}>
     =e^{iQ_x X}\Delta(\vec{Q}),
\end{equation}
where $\Delta(\vec{Q})$ is the order parameter. The cohesive
energy can be calculated in the same way as it has been done in
Refs.[\onlinecite{Yosh,Koulakov,Yang2}]:
\begin{equation}
E_{coh}=\frac{1}{2\nu}\sum_{\vec{Q}\neq 0}u_{HF}(\vec{Q})|\Delta(\vec{Q})|^2, \label{ground}
\end{equation}
where $\nu$ is the filling factor of the lowest Landau level.

We carry out the self-consistent HF computation on a hexagonal lattice with the wave vectors of the order
parameters as $\vec{Q}=[(j+{1\over 2 })Q_0,{\sqrt{3}\over 2}kQ_0]$, where $j$ and $k$ are integers. Following the
procedure in Ref.[\onlinecite{Yosh}], when $NQ_x^0Q_y^0l_B^2=2M\pi$, with $N$ and $M$ being integers, the Landau
level splits into $N$ Hofstadter bands. When $N=6$ and $M=1$ the WC has the lowest energy. In our calculations, we
choose $\nu=0.12$, at which the ground state is a Wigner crystal.

Figure 1 displays the dependence of the cohesive energy of the
electronic crystal on the tilted angle $\theta$ for various
orientation angles of the crystal to the in-plane magnetic field,
where $\phi$ is the angle between one side of the hexagonal
lattice with the in-plane field. It shows that of the two typical
configurations of orientation with respect to the in-plane field:
the $[100]$ and the $[110]$, the energy is always lower for
$[110]$ direction parallel to the in-plane field. The energy
difference increases with the tilted angle. In Fig.2 we plot the
relation of the cohesive energy of the hexagonal lattice with the
characteristic frequency $\Omega$ for tilted angles $\theta=0^0$
and $\theta=43.2^0$, respectively. The $[110]$ direction is along
the in-plane field for both curves. The smaller the characteristic
frequency is, i.e., the wider the quantum well is, the higher the
cohesive energy. The in-plane magnetic field can lower the
cohesive energy of the electronic crystal. From Fig.2 one can see
that the energy difference becomes larger for wider quantum wells.
Our calculations show that when the tilted angle increases
further, the energy difference will increase significantly,
implying that the in-plane magnetic field stabilizes the
orientation of the electronic crystal more effectively.

\section{transport property of the electronic solid}
Pinning of the Wigner crystal by impurities as a
result of breaking of the translational invariance has been widely
investigated\cite{Kuk,Millis}. Sherman\cite{Sher} had studied the
angular pinning and the domain structure of the electronic crystal
mediated by acoustic-phonon in $III-V$ semiconductor. Our
calculation shows that the in-plane magnetic field may serve as a
tunable means to probe the orientation of the crystal. Below we
will explore the implications of the preferred orientation of the
electronic crystal to the transport measurements. In realistic
samples a domain structure is formed due to a finite impurity
density. The electrons in each domain are ordered as they are in
the crystal. In the absence of the in-plane field, each domain
orients randomly, just like the domains in ferromagnets. An ideal
electronic crystal is an insulator and the conductivity
$\sigma_{xx}\propto e^{-\Delta_0/k_BT}$\cite{Chui,Li}. This
thermal activation form of the conductivity implies that the
electrons are hopping with a fixed range mechanism. It has been
confirmed by experiments that $\Delta_0$ is of the order 1K
\cite{jiang}. In a realistic domain structure, however, the
electrons may hop between the edges of the randomly oriented
domains. Since the experimental reachable temperature may be as
low as 10mK, the variable range hopping mechanism may work in this
temperature regime\cite{mott}. In the following, we will discuss
the possible consequence of the tilted field on the transport
properties.

In the usual Anderson localization the envelope of the wave
function falls off exponentially as $\phi_0\sim e^{-r/\xi}$, where
$\xi$ is the localization length. With a magnetic field the
electronic wave function of a perfect system is essentially a
Gaussian as $\phi_m\sim e^{-r^2/2l_B^2}$. In a slightly disordered
system one can think that some of the states will be pinned at
certain isolated impurity sites. The mixing of these states due to
quantum-mechanical tunnelling leads to a simple exponential tail
in the wave function\cite{Li}. In a strong magnetic field, the
electrons condense into a crystal at lower filling factors. When
the temperature is high enough the transport is of the thermal
activation form, which implies that the electrons are hopping with
a fixed range mechanism\cite{Chui,Li}. The hopping range is
determined by $R_0=\sqrt{1/\pi n_I}$, where $n_I$ is the impurity
density. However, localized states may exist along the edges of
the domains of the electronic crystal because of the impurities.
When the temperature is sufficiently low such that there is nearly
no phonon with energy to assist the electron making the nearest
hopping, Mott's variable hopping mechanism\cite{mott} allows the
electrons hop a larger distance $R>R_0$ to a state which has a
smaller energy difference $\Delta(R)$. In turn, the hopping
conduction is determined by the typical decay rate of the tails of
the wave function. The hopping probability is then given by
\begin{equation}
p\propto \exp[-R/\xi-\Delta/k_BT], \label{prb}
\end{equation}
where $R=|\vec R_i-\vec R_j|$ and $\Delta$ is the activation
energy.

As in the quantum Hall effect regime, strong interaction between
electrons leads to the system condensing into a WC. The Coulomb
gap depresses the density of states near the Fermi surface
\cite{efros2,YM}. Efros {\it et al} \cite{efros} had derived the
density of states near the Fermi surface $N(E)\propto |\Delta E|=
|E-E_F|$. From these considerations, one can get the conductivity
in the variable range hopping as\cite{efros2},
\begin{equation}
\sigma_{xx}\propto p\propto e^{-A/T^{1/2}},
\end{equation}
where $A=[\frac{4\hbar v_F}{k_B \xi}]^{1/2}$. The characteristic temperature $T_0$ above which the fixed range
hopping dominates is determined by $\bar R=R_0$, namely
\begin{eqnarray}
k_BT_0=\frac{2\hbar^2l_B}{m_b}(\pi n_I)^{3/2}.\label{t0}
\end{eqnarray}

Now, we discuss the possible effect of the tilted field. As we
have discussed, the existence of an in-plane field deforms the
electron wave function. However, this wave function deformation
does not qualitatively change the electron hopping mechanism at a
given temperature. The major effect of the tilted field would be
on the variation of $T_0$. As we have shown, the in-plane field
lowers the cohesive energy of the Wigner crystal and forces the
domains align to the same direction. Thus, the role of the
in-plane field is to integrate the domains into larger ones. In
this way, the in-plane field causes some of impurities to be
irrelevant and therefore reduces the effective impurity density.
To determine $T_0$ from (\ref{t0}), only the relevant impurities
should be counted in. Hence, one can replace $n_I$ by an effective
impurity density $n_I(B_\parallel)$. From Eq.(\ref{t0}), we see
that $T_0$ is sensitive to $n_I(B_\parallel)$. In a strong
magnetic field the decay length is comparable to the cyclotron
radius $\xi\sim R_c$\cite{Fogler}. For a sample with $n_I\sim
1.0\times 10^8$cm$^{-2}$, we estimate $T_0\sim 40$ mK. This
temperature is experimentally reachable. Therefore, it is possible
to observe a change of transport behavior that displays a
crossover from the variable to the fixed range hopping under
proper parameters and temperature as the tilted angle rises.

\section{summary}
We have shown that the WC has a preferred orientation with respect
to the in-plane magnetic field. We argued that there are domains
in a realistic sample and predicted that the temperature
dependence of the transport behavior may be different in different
temperature regime. Moreover, we emphasized that this preferred
orientation of the crystal may lead to an in-plane field induced
crossover from the variable range hopping to the fixed range
hopping of the transport mechanism in the 2D electronic solid. We
expect future experiments to verify our prediction.

This work was supported by a grant from Beijing Normal University.

\centerline {Figure Captions}

Figure 1 A 3D graph of the cohesive energy of the WC with respect to the tilted angle $\theta$ as well as the
orientation angle $\phi$.

Figure 2 The cohesive energy of the hexagonal lattice with the characteristic frequency $\Omega$ for tilted angle
$\theta=0^0$ and $\theta=43.2^0$, respectively. The energy difference increases with decreased $\Omega$.

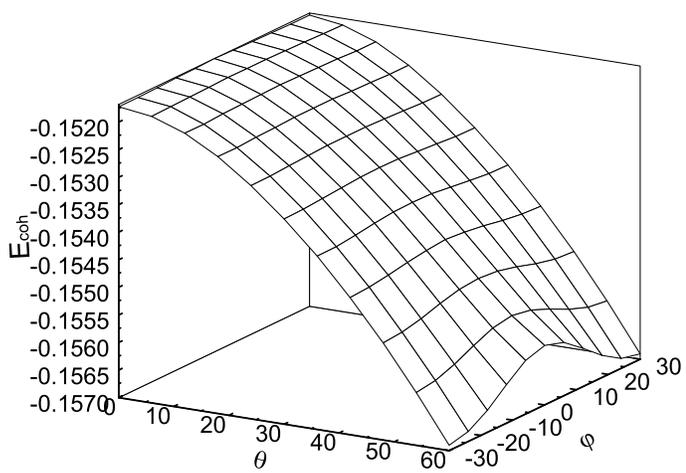
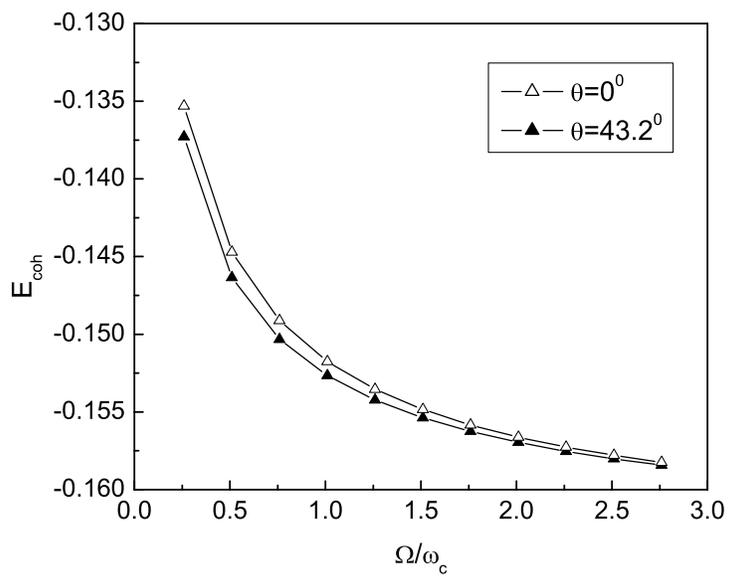
\end{document}